\documentclass[twocolumn,preprintnumbers,amsmath,amssymb]{revtex4}%,showkeys
\usepackage{graphicx}% Include figure files
\usepackage{amsmath}
\usepackage{subeqnarray}
\usepackage{cases}
\usepackage{amsmath,epsfig,amssymb,subfigure,bm,dsfont}
\usepackage{lipsum}
\usepackage{graphics}

\usepackage{subfigure}
\usepackage{caption}
\usepackage{overpic}
\begin{document}

\title{Enhanced mechanical squeezing in an optomechanical system via backward stimulated Brillouin scattering}

\author{Shan-Shan Chen $^{1}$, Na-Na Zhang $^{1}$, Yong-Rui Guo $^{1}$, Huan Yang $^{1,}$\footnote{yanghuan@cqupt.edu.cn} and Yong Ma$^{1,2,}$\footnote{mayong@cqupt.edu.cn}}

\address{$^{1}$School of Optoelectronic Engineering, Chongqing University of Posts and Telecommunications, Chongqing, 400065, China \\
$^{2}$China Electronics Technology Group Corporation 44th Research Institute, Chongqing, 400065, China }

\begin{abstract}
We investigate theoretically the enhancement of mechanical squeezing in a multimode optomechanical system by introducing a coherent phonon-photon interaction via the backward stimulated Brillouin scattering (BSBS) process. The coherent photon-phonon interaction where two optical modes couple to a Brillouin acoustic mode with a large decay rate provides an extra channel for the cooling of a Duffing mechanical oscillator. The squeezing degree and the robustness to the thermal noises of the Duffing mechanical mode can be enhanced greatly. When the Duffing nonlinearity is weak, the squeezing degree of the mechanical mode in the presence of BSBS can be improved more than one order of magnitude compared with the absence of BSBS. Our scheme may be extended to other quantum systems to study novel quantum effects.
\end{abstract}

%\pacs{03.67.Lx, 03.67.Ac, 03.67.Mn, 42.30.-d }

%\keywords{\emph{nonreciprocal mechanical squeezing, optomechanical system, Sagnac effect, Duffing nonlinearity}}

\maketitle

\section{Introduction}\label{sec1}
Stimulated Brillouin scattering (SBS) \cite {ShenPR1965,YarivIJQE1965} is an important nonlinear optical effect in the field of laser physics. SBS describes a coherent interaction between light and travelling acoustic wave, which originates from the electrostriction and photoelastic effects. This emerging subject leads to some remarkable and interesting topics, such as high coherence lasers \cite {HeNatPhys2020,LinAPL2014,HondaAPL2018,BaiPRL2021}, Brillouin gyroscope \cite {LiOptica2017} and Brillouin cooling \cite {BahlNatPhy2012}. The forward stimulated Brillouin scattering (FSBS) \cite {SavchenkovOL2011,BahlNatPhy2012,BahlNatCom2011,DongNatCommu2015} and the backward stimulated Brillouin scattering (BSBS) \cite {GrudininPRL2009,TomesPRL2009,EnzianOptica2019,EnzianPRL2021} are the research hotspots in recent years. In the BSBS, a Stokes optical signal is generated when a pump light photoelastically backscatters from a travelling acoustic wave. The remaining pump light and the Stokes optical signal electrostrictively excite the acoustic wave at the beat frequency of two optical signals. The wave vector conservation leads to the emergency of an acoustic mode at high frequency. In contrary to the BSBS, the Stokes optical signal in the FSBS copropagates with the forward-scattered pump light, and the wave vector conservation results in a much lower acoustic frequency. The higher resonance frequency and the larger mechanical decay rate \cite{GrudininPRL2009,TomesPRL2009,EnzianOptica2019,EnzianPRL2021} of the acoustic mode in the BSBS hold immense promise for the quantum control. First, a high-frequency acoustic mode enables low thermal occupation, the effect of the thermal noises on the quantum effect can be reduced. Second, a large mechanical decay rate is favorable for the control of the optical and mechanical quantum states, the related theoretical and experimental schemes have been reported in the quantum-limited amplification \cite{NunnenkampPRL2014}, the enhancement of the optomechanical entanglement and cooling \cite{ZhangPRA2020}, and the dissipatively controlled optomechanical interaction \cite{ShenPRL2021}. The proposal of these schemes open the door to the application of BSBS in the field of cavity optomechanics. Recently, the BSBS has been used to effectively enhance the optomechanical cooling and amplification of the target mechanical oscillations \cite{ShenPRL2021}, and the investigation of the quantum properties based on the optomechanical system (OMS) via the BSBS is attracting increasing attention.

The OMS \cite{KippenbergScience2008,AspelmeyerPhysToday2012,AspelmeyerRMP2014,ChenADP2018,ChenPRA2019} where optical field couples to mechanical oscillator via radiation pressure force is an important platform in the quantum information processing. The fundamental study of the OMS includes the optomechanically induced transparency \cite{HuangPRA2011,WeiScience2010,Safavi-NaeiniNature2011}, the mechanical squeezing \cite {LvPRA2015,LiaoPRA2011,ChenADP2020}, the ground-state cooling of the mechanical oscillator \cite{TeufelNature2011,MeenehanPRX2015,ClarkNature2017}, and so on. The phonon-photon interaction in the OMS arises from the radiation pressure force, while the electrostrictive force determines the generation of the phonon-photon interaction in the BSBS. The combination of the cavity optomechanics and the BSBS provides new opportunities for the quantum control of the phonon at the quantum level. The quantum squeezing of the mechanical mode has been at the center of many important developments in cavity optomechanics due to its potential applications in improving the precision of quantum measurement and the continuous-variable quantum-information processing \cite{CavesRMP1980,LaHayeScience2004,BraunsteinRMP2005}. The mechanical squeezing based on the OMS can be achievable via many methods, including
the squeezing transfer between a cavity driven by a squeezed optical field and a mechanical mode \cite{JahnePRA2009,HuangPRA2010}, the reservoir engineering technique \cite{KronwaldPRA2013,WollmanScience2015,ZhangPRA2019}, and the periodically amplitude modulation of the driving laser \cite{MariPRL2009,LiaoPRA2011,SchmidtNJP2012,BaiPhotRe2019}. The ground-state cooling of the mechanical oscillator is the premise of the generation of the mechanical squeezing. With the proposal of cooling the mechanical oscillator by using a Brillouin acoustic mode with high frequency and large decay rate in the BSBS \cite{ShenPRL2021}, BSBS is expected to be used for the preparation of a strong quantum squeezing of the mechanical mode.

In this paper, we propose a scheme to enhance the quantum squeezing of the mechanical mode based on an OMS via the BSBS. We find that, when the effective frequency detuning of the Brillouin acoustic mode and the phonon-photon coupling via the BSBS process are chosen properly, the introduction of the phonon-photon interaction via the BSBS process provides a cooling channel for the target mechanical oscillator, the squeezing degree of the mechanical mode can be improved effectively. For a weak Duffing nonlinearity, the squeezing degree of the mechanical mode can be improved more than one order of magnitude due to BSBS, a strong mechanical squeezing which surpasses the 3 dB limit can be achieved. Our scheme provides a good theoretical guidance for studying more interesting quantum effects based on an optomechanical cavity.

This paper is organized as follows. In Sect.~\ref{sec2}, we describe the multimode optomechnaical model and obtain the linearized Hamiltonian. We study the effect of the phonon-photon interaction via the BSBS process on the steady-state position variance of the mechanical mode in Sect.~\ref{sec3}, and discuss the feasibility of the experiment in Sec.~\ref{sec4}. A summary is given in Sec.~\ref{sec5}.

\section{Theoretical model}\label{sec2}

 We consider a multimode OMS in which two optical modes $a_1,a_2$ couple to a travelling acoustic mode $b$ via the BSBS process, and a Duffing mechanical oscillator $c$ with nonlinear amplitude $\eta$ couples to two optical modes $a_{1},a_{2}$ simultaneously via the radiation pressure force. The Hamiltonian of the full system can be described as ($\hbar=1$)
\begin{eqnarray} \label{eq1}
H_{\rm{T}}=H_{0}+H_{\rm{BSBS}}+H_{\rm{OM}}+H_{\rm{D}}+H_{\rm{F}},
\end{eqnarray}

where
\begin{eqnarray} \label{eq2}
H_{0}&=&\sum\limits_{j}\omega_{j}a_{j}^{\dagger}a_j+\omega_{b}b^\dagger b+\omega_mc^\dagger c,\nonumber\\
H_{\rm{BSBS}}&=&-g_b(a_1^\dagger a_2b+a_1a_2^\dagger b^\dagger),\nonumber\\
H_{\rm{OM}}&=&-\sum\limits_{j}g_{c_j}a_{j}^{\dagger}a_j(c+c^\dagger),\nonumber\\
H_{\rm{D}}&=&i\sum\limits_{j}\sqrt{\kappa_{ex_j}\varepsilon_{d_j}}(a_j^\dagger e^{-i\omega_{d_j}t}-a_je^{i\omega_{d_j}t}),\nonumber\\
H_{\rm{DN}}&=&\frac{\eta}{2}(c+c^\dagger)^4.\nonumber\\
\end{eqnarray}
Here $H_0$ is the free Hamiltonian of the system. $a_j$ ($a_j^\dagger$) for $j=1,2$, $b$ ($b^\dagger$) and $c$ ($c^\dagger$) are the annihilation (creation) operators of the optical mode with frequency $\omega_j$, the Brillouin acoustic mode with frequency $\omega_b$ and the mechanical mode with frequency $\omega_m$, respectively. $H_{\rm{BSBS}}$ is the Hamiltonian of a triply resonant phonon-photon interaction via the BSBS process, $g_b$ is the single-photon Brillouin coupling strength between two optical modes $a_1$, $a_2$ and a Brillouin acoustic mode $b$. $H_{\rm{OM}}$ describes the radiation-pressure coupling between the optical mode $a_{1(2)}$ and the target mechanical mode $c$ with a single-photon coupling strength $g_{c_{1(2)}}$. $H_{\rm{D}}$ is the driving term of the optical mode $a_{1(2)}$ with total decay rate $\kappa_{1(2)}=\kappa_{0_{1(2)}}+\kappa_{ex_{1(2)}}$ which includes an intrinsic decay rate $\kappa_{0_{1(2)}}$ and an external decay rate $\kappa_{ex_{1(2)}}$. $\varepsilon_{d_{1(2)}}$ and $\omega_{d_{1(2)}}$ are amplitude and frequency of the external control field, respectively. $H_{\rm{DN}}$ is the Duffing nonlinear term, the strong nonlinearity can be realized by the coupling between a mechanical mode and an ancilla system \cite{BlencowePR2004,XiangRMP2013,JacobsPRL2009,TianPRB2001}.

\begin{figure}[!ht]
\begin{center}
\includegraphics[width=6.5cm,angle=0]{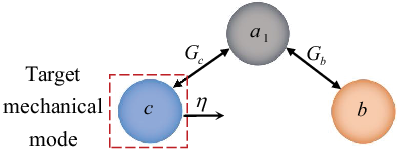}
\caption{Schematic of the effective three-mode optomechanical device which consists of an optical mode $a_1$, an acoustic mode $b$ and a mechanical mode $c$ (nonlinearity $\eta$). $G_c$ ($G_b$) is the effective phonon-photon coupling between the mechanical (acoustic) mode $c$ ($b$) and the optical mode $a_1$, which arises from the radiation pressure (electrostrictive) force.}\label{fig1}
\end{center}
\end{figure}

When the optical mode $a_2$ is pumped by a very strong control field and the amplitude of the Brillouin acoustic mode $b$ is very weak, this control field can be treated classically. In the weak single-photon Brillouin coupling regime $g_b\ll\omega_b,\kappa_{1(2)},$ and the duration of control pulse satisfies $\tau_p\gg1/\kappa_{2}$, the steady state solution $\alpha_2$ of the optical mode $a_2$ can be obtained as
\begin{eqnarray} \label{eq3}
\alpha_2=\frac{\sqrt{\kappa_{ex_{2}}}\varepsilon_{d_2}}{\kappa_{2}/2+i\Delta_{2}},
\end{eqnarray}
where $\Delta_2=\omega_2-\omega_{d_2}$. In the rotation frame with $H_r=\omega_{d_1}a_{1}^{\dagger}a_1+\omega_{d_2}a_{2}^{\dagger}a_2+(\omega_{d_1}-\omega_{d_2})b^\dagger b$, the Hamiltonian Eq. (\ref{eq2}) of the system can be linearized as

\begin{eqnarray} \label{eq4}
H_{lin}&=&\Delta_1a_{1}^{\dagger}a_1+\Delta_{b}b^\dagger b+\tilde{\omega}_mc^\dagger c\nonumber\\
&-&[G_{b}a_1^\dagger b+G_{c}a_1^\dagger(c^\dagger+c)+\text{H.c.}]\nonumber\\
&+&\Lambda(c^2+c^{\dagger2}),
\end{eqnarray}

where
\begin{eqnarray} \label{eq5}
&&\Delta_1=\omega_1-\omega_{d_1}-2g_{c_1}\beta, \ \ \ \ \Delta_b=\omega_b+\omega_{d_2}-\omega_{d_1},\nonumber\\
&&\tilde{\omega}_m=\omega_m+2\Lambda,\ \ \ \ \ \ \ \ \ \ \  \ \ G_{b}=g_b\alpha_2,\nonumber\\
&&G_{c}=g_{c_1}\alpha_1,\ \ \ \ \ \ \ \ \ \ \ \ \ \ \ \ \ \ \ \Lambda=3\eta(4\beta^2+1),\nonumber\\
\end{eqnarray}
here we only retain the linear and bilinear terms due to $g_{c_1}, \eta\beta\ll\Lambda,G_{c}$, and have used the steady-state amplitudes $\alpha_1$ and $\beta$ of optical mode $a_1$ and mechanical mode $c$, respectively, which fulfill
\begin{eqnarray} \label{eq6}
&&\alpha_1=\frac{\sqrt{\kappa_{ex_{1}}}\varepsilon_{d_1}}{\kappa_{1}/2+i\Delta_{1}},\nonumber\\
&&16\eta\beta^3+(12\eta+\omega_m)\beta-g_{c_1}|\alpha_1|^2=0,
\end{eqnarray}
where the mechanical decay rate $\gamma_m$ of the mechanical mode $c$ is so weak $\gamma_m\ll\kappa_{1(2)}$ that the $\gamma_m$-dependent terms have been dropped. Hence, the effective optomechanical model used in our scheme is equivalent to a three-mode optomechanical device, as shwon in Fig. ~\ref{fig1}.

\section{The steady-state position fluctuation of the mechanical mode}\label{sec3}

In order to realize the steady-state mechanical squeezing of the target mechanical mode $c$, we introduce the squeezing transformation $c=\cosh(r)c_s-\sinh(r)c_s^\dagger$ with
\begin{eqnarray} \label{eq7}
r=\frac{1}{4}\ln(1+4\Lambda/\omega_m)
\end{eqnarray}
to the Hamiltonian Eq. (\ref{eq4}), where $c_s$ is the squeezed mechanical mode, the effective Hamiltonian $H_{lin}$ of the system in the squeezing transformation frame is given by

\begin{eqnarray} \label{eq8}
H_{eff}&=&\Delta_1a_{1}^{\dagger}a_1+\Delta_{b}b^\dagger b+\omega'_mc_s^\dagger c_s\nonumber\\
&-&[G_{b}a_1^\dagger b+G'_{c}a_1^\dagger(c_s^\dagger+c_s)+\text{H.c.}],\nonumber\\
\end{eqnarray}
where $\omega'_m=\omega_m\sqrt{1+4\Lambda/\omega_m}$ and $\ G'_{c}=G_{c}(1+4\Lambda/\omega_m)^{-\frac{1}{4}}$ are the effective mechanical frequency and the optomechanical coupling in the squeezing transformation frame, respectively. The dynamics for the optical mode $a_1$, the acoustic mode $b$ and the squeezed mechanical mode $c_s$ can be described by the quantum Langevin equations
\begin{eqnarray} \label{eq9}
\dot{a}_1&=&-(\frac{\kappa_1}{2}+i\Delta_1)a_1+iG_{b}b+iG'_{c}(c_s^\dagger+c_s)+\sqrt{\kappa_1}a_1^{in}(t),\nonumber\\
\dot{b}&=&-(\frac{\gamma_b}{2}+i\Delta_b)b+iG_{b}^*a_1+\sqrt{\gamma_b}b^{in}(t),\nonumber\\
\dot{c}_s&=&-(\frac{\gamma_m}{2}+i\omega'_m)c_s+i(G'_{c}a_1^\dagger+G_{c}^{'*}a_1)+\sqrt{\gamma_m}c_s^{in}(t),\nonumber\\
\end{eqnarray}
where $\gamma_b$ is the decay rate of the Brillouin acoustic mode. The input vacuum noise $a_1^{in}(t)$ for the cavity mode and the thermal noises $b^{in}(t),c_s^{in}(t)$ for the mechanical modes satisfy

\begin{eqnarray} \label{eq10}
\langle a_1^{in}(t)a_1^{in\dagger}(t')\rangle&=&\langle b^{in}(t)b^{in\dagger}(t')\rangle=\delta(t-t'),\nonumber\\
\langle c_s^{in\dagger}(t)c_s^{in}(t')\rangle&=&N_{eff}\delta(t-t'),\nonumber\\
\langle c_s^{in}(t)c_s^{in}(t')\rangle&=&M_{eff}\delta(t-t'),\nonumber\\
\end{eqnarray}
where $N_{eff}=\cosh(2r)n_m+\sinh^2(r)$ and $M_{eff}=\sinh(2r)(n_m+1/2)$, $n_m=[\exp(\hbar\omega_m/k_bT)-1]^{-1}$ is the equilibrium phonon occupation, $k_b$ is the Boltzmann constant, and we have neglected the thermal noise of the high-frequency Brillouin mode $b$ due to $\omega_b\gg\omega_m$.
To calculate the position variance of the mechanical mode $c$, we define the column vectors which contain the amplitude and phase quadrature operators $X_o=(o+o^\dagger)/\sqrt{2}, P_o=i(o^\dagger-o)/\sqrt{2}$
and the corresponding noise quadrature operators $X_o^{in}=(o^{in}+o^{\dagger in})/\sqrt{2}, P_o^{in}=i(o^{in\dagger}-o^{in})/\sqrt{2}$ ($o=a_1,b,c_s$) for the fluctuations $D=[X_{a_1},P_{a_1},X_b,P_b,X_{c_s},P_{c_s}]^T$ and the quantum noises $D^{in}=[X_{a_{1}}^{in},P_{a_{1}}^{in},X_b^{in},P_b^{in},X_{c_s}^{in},P_{c_s}^{in}]^T$. Then the evolution of the system in Eq. (\ref{eq9}) can be formulated as

\begin{eqnarray} \label{eq11}
\frac{dD}{dt}=MD+D^{in}
\end{eqnarray}
with a $6\times6$ matrix

\begin{eqnarray} \label{eq12}
M=\left[\begin{array}{cccccc}
    -\frac{\kappa_1}{2} & \Delta_1 &0 &-G_{b} &0 &0\\
    -\Delta_1 & -\frac{\kappa_1}{2} &G_{b} &0 &2G'_{c} & 0\\
    0 & -G_{b} &-\frac{\gamma_b}{2} &\Delta_b &0 & 0\\
    G_{b} &0 &-\Delta_b &-\frac{\gamma_b}{2}  &0 &0\\
    0 &0 &0 &0  &-\frac{\gamma_m}{2} &\omega'_m\\
   2G'_{c} &0 &0 &0  &-\omega'_m &\frac{\gamma_m}{2}\\
  \end{array}\right].
\end{eqnarray}
Here the coupling constans $G_{b}$ and $G'_{c}$ have been chosen as real numbers for simplicity. The stability of the system can be ensured when all the eigenvalues of $M$ have negative real parts. We have checked carefully the stability of the system in the following calculation.

The linearized Hamiltonian and the zero-mean Gaussian nature of the quantum noises ensure the Gaussian characteristics of the system, the time evolution of the system can be described by a $6\times6$ covariance matrix $V$ with elements $V_{kl}=\langle D_kD_l+D_lD_k\rangle/2$ ($k,l=1,2,3,4,5,6$), the motion equations in Eq. (\ref{eq11}) can be expressed as
\begin{eqnarray} \label{eq13}
\frac{dV}{dt}=MV+VM^T+A,
\end{eqnarray}
where the diagonal noise correlation matrix $A=\text{Diag}[\frac{\kappa_1}{2},\frac{\kappa_1}{2},\frac{\gamma_b}{2},\frac{\gamma_b}{2},\frac{e^{2r}\gamma_m(2n_m+1)}{2},\frac{e^{-2r}\gamma_m(2n_m+1)}{2}]$. When the system reaches a steady state for $t\rightarrow\infty$, we obtain a
Lyapunov equation
\begin{eqnarray} \label{eq14}
MV(\infty)+V(\infty)M^T=-A,
\end{eqnarray}
the steady-state position variance $\langle \delta X_{c}^2\rangle$ of the mechanical mode $c$ in the original frame can be obtained as
\begin{eqnarray} \label{eq15}
\langle \delta X_{c}^2\rangle=e^{-2r}V_{55},
\end{eqnarray}
the corresponding effective phonon number $n_{eff}$ of the mechanical oscillator $c$ in the original frame reads
\begin{eqnarray} \label{eq16}
n_{eff}=(e^{-2r}V_{55}+e^{2r}V_{66}-1)/2.
\end{eqnarray}
\begin{figure}
\centering
\subfigure{
\begin{minipage}[b]{0.5\textwidth}
\label{fig2a}%\includegraphics[width=6.0cm,angle=0]{fig2a.eps}
\begin{overpic}[width=6.0cm,angle=0]{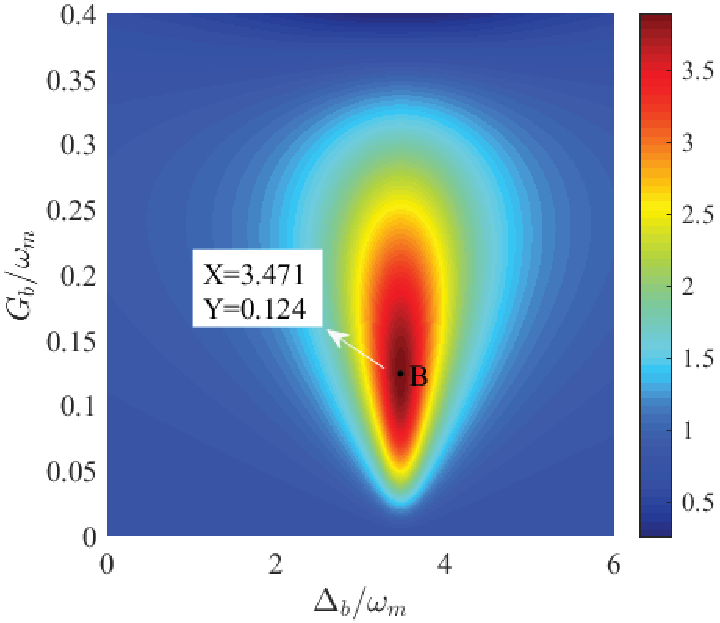}\put(0,83){(a)}\end{overpic}
%\subcaption*{fig2a}
\end{minipage}
}
\subfigure{
\begin{minipage}[b]{0.5\textwidth}
\label{fig2b}%\includegraphics[width=6.0cm,angle=0]{fig2b.eps}
\begin{overpic}[width=6.0cm,angle=0]{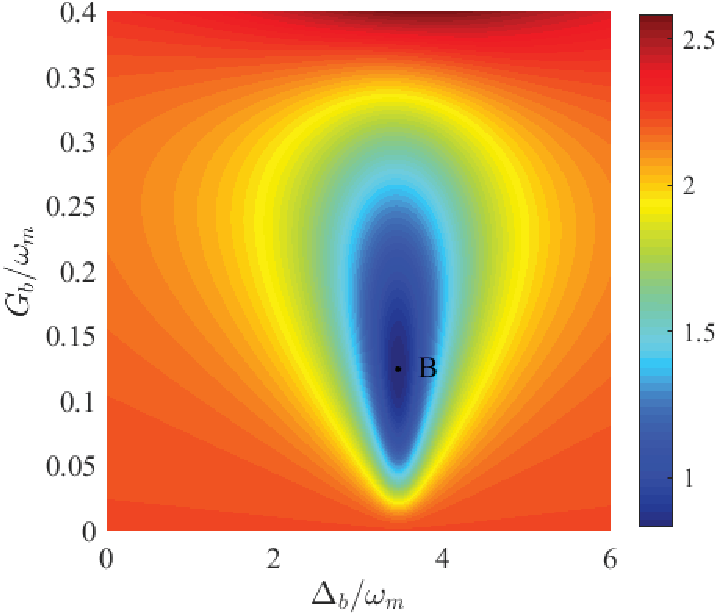}\put(0,83){(b)}\end{overpic}
\end{minipage}
}
\caption{Position variance $\langle \delta X_{c}^2\rangle$(dB) (a) and effective phonon number $n_{eff}$ (b) of the mechanical mode $c$ as a function of effective coupling $G_{b}$ and frequency detuning $\Delta_b$. Here we choose $\omega_m/(2\pi)=1$ MHz, $g_{c}=10^{-4}\omega_m$, $\kappa_1=0.02\omega_m$, $\gamma_b=0.4\omega_m$, $\gamma_m=10^{-4}\omega_m$, $\eta=10^{-4}\omega_m$, $n_m=100$, $G_{c}=0.15\omega_m$, $\Delta_1=\omega'_m$.} \label{fig2}
\end{figure}

\begin{figure}
\centering
\subfigure{
\begin{minipage}[b]{0.5\textwidth}
\label{fig3a}\begin{overpic}[width=6.0cm,angle=0]{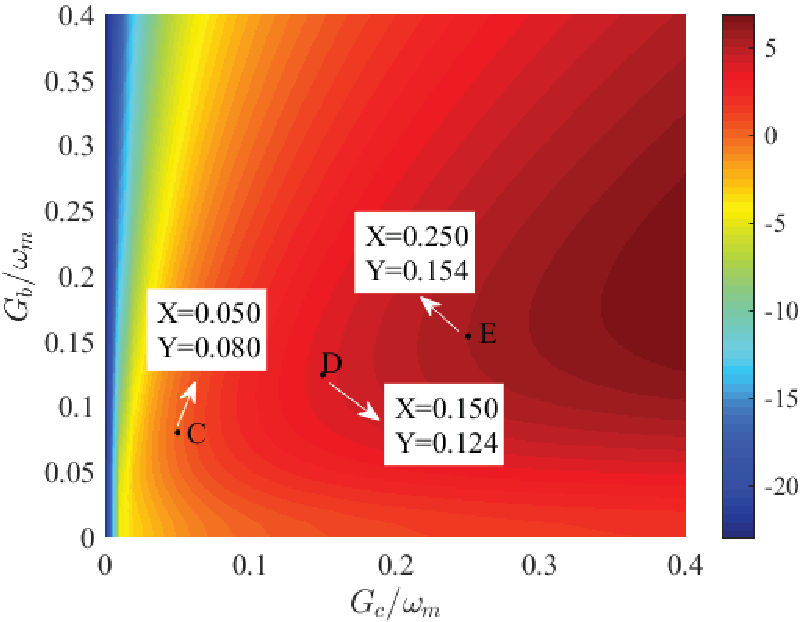}\put(-1,75){(a)}\end{overpic}
\end{minipage}
}
\subfigure{
\begin{minipage}[b]{0.5\textwidth}
\label{fig3b}\begin{overpic}[width=5.7cm,angle=0]{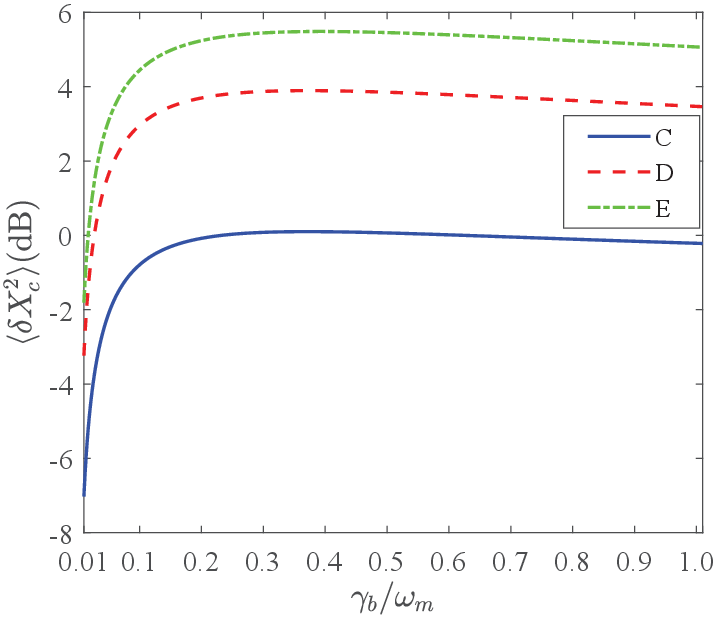}\put(-1.5,83){(b)}\end{overpic}
\end{minipage}
}
 \caption{(a) Position variance $\langle \delta X_{c}^2\rangle$(dB) of the mechanical mode $c$ versus the optomechanical coupling $G_{c}$ and the effective phonon-phonton coupling $G_{b}$ via the BSBS process. (b) Position variance $\langle \delta X_{c}^2\rangle$ of the mechanical mode versus the Brillouin acoustic damping rate $\gamma_b$ at the points C, D and E in Fig.~\ref{fig3a}. Here we choose $\Delta_b=\Delta_1=\omega'_m$, other parameters are the same as Fig.~\ref{fig2}.} \label{fig3}
\end{figure}

To evaluate the squeezing degree of the mechanical mode, we adopt the physical quantity $\langle \delta X_{c}^2\rangle\text{(dB)}=-10\log_{10}\frac{\langle \delta X_{c}^2\rangle}{\langle \delta X_{c}^2\rangle_{vac}}$ in the following calculations, where $\langle \delta X_{c}^2\rangle_{vac}=\frac{1}{2}$ is the position variance of the mechanical oscillator in the ground state. From now on, we study the effect of the phonon-photon interaction via the BSBS process on the position variance $\langle \delta X_{c}^2\rangle\text{(dB)}$ of the mechanical mode $c$. Here we choose the red-detuned resonance $\Delta_1=\omega'_m$ to cool the squeezed mechanical mode $c_s$ by suppressing the heating terms $G'_{c}(a_1^\dagger c_s^\dagger+a_1c_s)$. As shown in Fig. ~\ref{fig2a}, we plot the effects of the frequency detuning $\Delta_b$ and the effective phonon-photon coupling $G_{b}$ on the position variance $\langle \delta X_{c}^2\rangle\text{(dB)}$ of the mechanical mode $c$. The strongest squeezing appears at the resonant point $B$ ($\Delta_b=3.471\omega_m=\Delta_1$ and $G_b=0.124\omega_m$), because there is a most effective beamsplitter-type coupling at this resonance point, the BSBS is an anti-Stokes process, then the introduction of the phonon-photon interaction via the BSBS process provides an extra loss channel to cool the OMS \cite{ZhangPRA2020}, which can be seen clearly from Fig. ~\ref{fig2b}, a minimum effective phonon number $n_{eff}$ appears at point B, an optimal cooling effect can be obtained. Cooling is essential for the prepartion of the mechanical squeezing, so an optimal squeezing is achieved at the point B. It is clear that the effective coupling $G_{b}$ also has obvious influence on the position variance of the mechanical mode, an optimal squeezing can be obtained when a proper $G_b$ is chosen. Additionally, the optimal choice of $G_b$ is related to the effective optomechanical coupling $G_c$, which can be further depicted in Fig. ~\ref{fig3a}.

In Fig. ~\ref{fig3a}, we plot the position variance $\langle \delta X_{c}^2\rangle\text{(dB)}$ of the mechanical mode $c$ versus the couplings $G_{c}$ and $G_{b}$ when $\Delta_b=\Delta_1=\omega'_m$. It is clear that the position squeezing is firstly enhanced and then weakened with the increasing of the effective phonon-photon coupling $G_{b}$ for
a fixed $G_{c}$. For a given $G_{b}$, the squeezing degree of the mechanical mode is proportional to the optomechanical coupling $G_{c}$. The physical mechanics can be explained as follows. When $G_{b}$ is not very strong, the optomechanical interaction $G'_c$ between the optical mode $a_1$ and the squeezed mechanical mode $c_s$ can be neglected in the weak-coupling regime $G_{c}\ll0.01\omega_m$, then the effective Hamiltonian is expressed as $H'_{eff}=\Delta_{1}a_{1}^{\dagger}a_{1}+\Delta_{b}b^\dagger b-G_{b}(a_1b^\dagger+a_1^\dagger b)$, the fluctuation spectrum $S_{FF}(\omega)=\int dte^{i\omega t}\langle F(t)F(0)\rangle$ of the radiation-pressure force $F=a_1^\dagger+a_1$ can be derived as
\begin{eqnarray} \label{eq17}
S_{FF}(\omega)=\frac{\frac{|G_{b}|^2\gamma_b}{|\frac{\gamma_b}{2}-i(\omega-\triangle_b)|^2}+\kappa_1}{|\frac{\kappa_1}{2}-i(\omega-\triangle_1)+\frac{|G_{b}|^2}{\frac{\gamma_b}{2}-i(\omega-\Delta_b)}|^2}.
\end{eqnarray}
According to the rate equation \cite{AspelmeyerRMP2014,GuoPRA2014}, we obtain the final effective phonon number $n_{eff}^s$ of the mechanical oscillator $c_s$ in the squeezing frame as

\begin{eqnarray} \label{eq18}
&&n_{eff}^s=\langle c_s^\dagger c_s\rangle=\frac{\gamma_mN_{eff}+\gamma_cn_c}{\gamma_m+\gamma_c}\approx\frac{\gamma_mN_{eff}}{\gamma_m+\gamma_c},
\end{eqnarray}

with
\begin{eqnarray} \label{eq19}
\gamma_c&=&G^{'2}_{c}[S_{FF}(+\omega'_m)-S_{FF}(-\omega'_m)]\approx\frac{4G^{'2}_{c}}{\kappa_{eff}},\nonumber\\
\kappa_{eff}&=&\kappa_1+\frac{4|G_{b}|^2}{\gamma_b},\nonumber\\
n_c&=&\frac{S_{FF}(-\omega'_m)}{S_{FF}(+\omega'_m)-S_{FF}(-\omega'_m)},
\end{eqnarray}
and, $n_{eff}=\langle c^\dagger c\rangle$, in the original frame

\begin{eqnarray} \label{eq20}
n_{eff}=\cosh(2r)n_{eff}^s+\sinh^2(r),
\end{eqnarray}
under the resonance condition $\Delta_1=\Delta_b=\omega'_m$. Here we have used the approximations $\gamma_m\ll\kappa_1\ll\gamma_b\ll\omega'_m$ and $S_{FF}(-\omega'_m)\ll S_{FF}(\omega'_m)$. The corresponding position variance of the mechanical mode in the original frame can be calculated as
\begin{eqnarray} \label{eq21}
\langle \delta X_c^2\rangle&=&(n_{eff}^s+\frac{1}{2})e^{-2r}.
\end{eqnarray}
From Eqs. (\ref{eq18}) and Eq. (\ref{eq21}), it is clear that the linewidth $\kappa_{eff}$ of optical mode $a_1$ is broadened with the increasing of the effective phonon-photon coupling $G_{b}$ via the BSBS process in the weak coupling regime, which leads to the increase of the final phonon number $n_{eff}$, the ground-state cooling effect of the mechanical motion becomes worse, so the squeezing degree of the position of the mechanical mode $c$ decreases. Interestingly, when we enter into a large optomechanical coupling region $G_{c}\geq0.01\omega_m$ and $G_{b}$ is not very strong, the above analysis is no longer valid, the mechanical squeezing is enhanced greatly with increasing $G_b$. However, a very strong $G_{b}$ is harmful for the enhancement of the mechanical squeezing, which originates from the emergence of the supermodes $a_+=(a_1+b)/\sqrt{2}$ with frequency $\omega_{a_+}=\Delta_1-|G_{b}|$ and $a_-=(b-a_1)/\sqrt{2}$ with frequency $\omega_{a_-}=\Delta_1+|G_{b}|$, the interaction term $G'_{c}(a_+^\dagger c_s^\dagger+c_sa_+)$ cannot be omitted. For a given $G_{b}$, the increase of the weak optomechanical coupling $G_{c}$ leads to a larger $\gamma_c$ in Eq. (\ref{eq18}), a larger cooling rate is obtained, the squeezing degree of the mechanical mode is improved rapidly at the beginning. Whereas the non-resonant term $G'_{c}(a_1c_s+a_1^\dagger c_s^\dagger)$ cannot be ignored when the coupling $G_{c}$ is large enough, the cooling effect is bad again, so the squeezing degree of the mechanical mode increases more and more slowly. Based on the above analysis, we conclude that an optimal mechanical squeezing is achievable when the optomechanical coupling $G_{c}$ and the effective phonon-photon coupling $G_{b}$ caused by the BSBS are chosen properly, for example, at the points C ($G_c=0.050\omega_m,G_b=0.008\omega_m$), D ($G_c=0.150\omega_m,G_b=0.124\omega_m$) and E ($G_c=0.250\omega_m,G_b=0.154\omega_m$), the optimal mechanical squeezing degrees are 0.099, 3.892 and 5.486 respectively.

In the BSBS, the Brillouin acoustic mode has a large decay rate, which is a key factor in our scheme. In the above discussions, we have used the condition $\gamma_b\gg\kappa_1$.  To confirm the necessity of this condition, as shown in Fig.~\ref{fig3b}, we plot the position variance $\langle \delta X_{c}^2\rangle$ of the mechanical mode versus the Brillouin acoustic decay rate $\gamma_b$ at the points C, D and E in Fig.~\ref{fig3a}. Obviously, the squeezing degree grows up rapidly to a maximum and then decreases slightly as the acoustic damping rate $\gamma_b$ increases. The strongest squeezing at the points C, D and E all appear in the region where $\gamma_b\gg\kappa_1=0.02\omega_m$.

\begin{figure}[!ht]
\begin{center}
\includegraphics[width=6.5cm,angle=0]{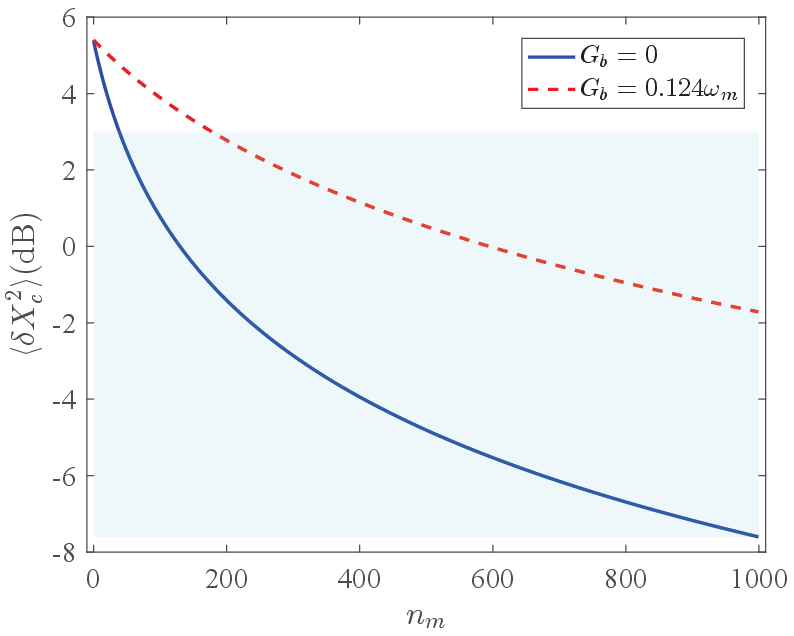}
\caption{Position variance $\langle \delta X_{c}^2\rangle$ of the mechanical mode $c$ as a function of the thermal phonon number $n_m$ for $G_{b}=0$ (black solid line) and $G_{b}=0.124\omega_m$ (red dashed line). Here we choose $\Delta_b=\Delta_1=\omega'_m$, other parameters are the same as in Fig.~\ref{fig2}. The shadowed blue bottom region corresponds to squeezing below the 3 dB limit.}\label{fig4}
\end{center}
\end{figure}

To examine the effect of the mechanical thermal noises, we plot the position variance $\langle \delta X_{c}^2\rangle\text{(dB)}$ of the mechanical mode as a function of the thermal phonon number $n_m$ in Fig. ~\ref{fig4}. Intuitively, the thermal phonon number has a negative influence on the generation of the mechanical squeezing. When $n_m\neq0$, the introduction of the effective phonon-photon coupling $G_{b}$ (red dashed line) contributes to the cooling of the target mechanical oscillator, the mechanical squeezing is enhanced greatly, the robustness of the mechanical squeezing to the thermal noise of the mechanical environment can be greatly improved compared with the case without the effective coupling $G_{b}=0$ (blue solid line).

\begin{figure}[!ht]
\begin{center}
\includegraphics[width=6.5cm,angle=0]{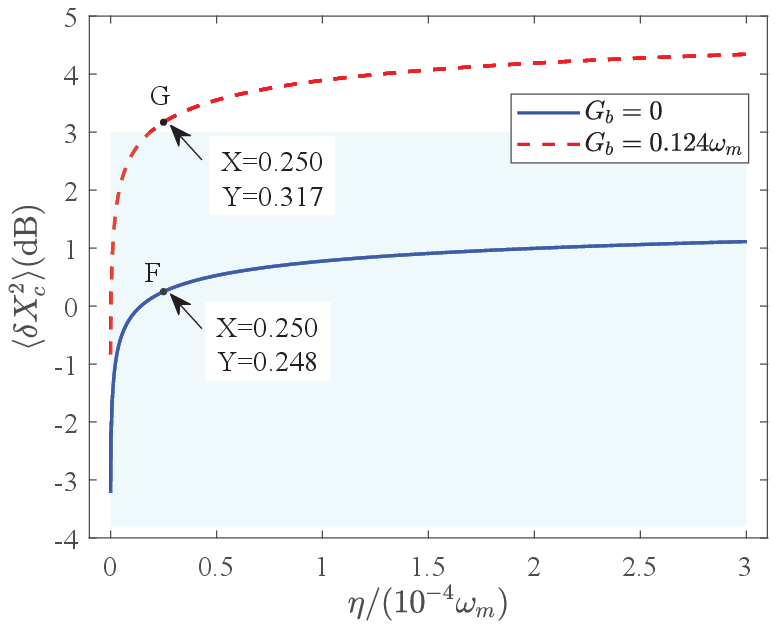}
\caption{Position variance $\langle \delta X_{c}^2\rangle$ of the mechanical mode $c$ as a function of the Duffing nonlinearity amplitude $\eta$ for $G_{b}=0$ (blue solid line) and $G_{b}=0.124\omega_m$ (red dashed line). Here we choose $\Delta_b=\Delta_1=\omega'_m$, other parameters are the same as in Fig.~\ref{fig2}. The shadowed blue bottom region corresponds to squeezing below the 3 dB limit.}\label{fig5}
\end{center}
\end{figure}

In Fig. ~\ref{fig5}, we plot the position variance $\langle \delta X_{c}^2\rangle\text{(dB)}$ of the mechanical mode $c$ as a function of the Duffing nonlinearity amplitude $\eta$ for $G_b=0$ (blue solid line) and $G_{b}=0.124\omega_m$ (red dashed line). As seen, the Duffing nonlinearity is necessary for the generation of the mechanical squeezing. The squeezing degree can be further improved when a proper effective phonon-photon coupling caused by the BSBS such as $G_{b}=0.15\omega_m$ is introduced compared to the case without the effective couping $G_b=0$. For example, for a weak nonlinearity $\eta=0.25\times10^{-4}\omega_m$, the squeezing degree $\langle \delta X_{c}^2\rangle=3.17$ dB at point G still exceeds the 3 dB limit (as the red line above the shadowed blue region shows) in the case of $G_b=0.124\omega_m$, which is about 13 times than that $\langle \delta X_{c}^2\rangle=0.248$ dB at point F when $G_b=0$.
\section{discussion} \label{sec4}

Our optomechanical model can be realized based on the advanced experimental technology. First, the triply resonant phonon-photon interaction via the BSBS process has been achieved experimentally in the platform of whispering-gallery-mode OMS \cite{ShenPRL2021}, and the frequency of the Brillouin acoustic mode has reached about 10 GHZ \cite{ShenPRL2021}, which indicates that it is reasonable to ignore the thermal noises of the acoustic mode, and the Brillouin acoustic mode with large decay rate has been proved to be effective for the cooling of the target mechanical oscillator \cite{ZhangPRA2020}.  Second, the preparation of the Duffing nonlinearity has been investigated in Refs. \cite{BlencowePR2004,XiangRMP2013,JacobsPRL2009,TianPRB2001}, the amplitude $\eta=10^{-4}\omega_m$ is achievable \cite{LvPRA2015}. Third, to reduce the negative effect of the quantum noises, it is necessary to precool the system \cite{ParkNatPhys2009}, and the related measurements can be performed via the dilution refrigeration or the sideband cooling cooling \cite{ParkNatPhys2009,SchliesserNatPhy2009,ChanNature2011,TeufelNature2011}.
Finally, the mechanical squeezing can be detected though the homodyne detection technology.

\section{summary} \label{sec5}

We have proposed an optomechanical scheme for preparing a strong steady-state mechanical squeezing via BSBS. When the effective frequency detuning of the Brillouin acoustic mode equals to that of the optical mode, and the optomechanical coupling and the phonon-photon coupling via the BSBS process are selected reasonably, the introduction of the phonon-photon interaction enables the enhancement of the mechanical squeezing under the red-detuning resonant condition in the squeezing transformation frame. In the weak optomechanical coupling regime, the increase of the effective phonon-photon coupling caused by the BSBS broads the linewidth of the optical mode, the cooling efficient of the mechanical mode decreases, the mechanical squeezing becomes worse. However, for a strong optomechanical coupling, the mechanical squeezing can be enhanced effectively in the presence of the phonon-photon coupling via the BSBS process. But the very strong phonon-photon coupling leads to the emergence of the supermodes, which is harmful for the phonon cooling, the squeezing degree reduces. When the Duffing nonlinearity is weak, compared to the case without the BSBS, the mechanical squeezing prepared in the presence of the BSBS is more robust to the quantum noises, and the squeezing degree can be improved by more than an order of magnitude.

\section*{ACKNOWLEDGEMENTS}

This work was supported by the Natural Science Foundation of Chongqing CSTC under Grants No. CSTB2022NSCQ-BHX0020, the China Electronics Technology Group Corporation 44th Research Institute Grants No.6310001-2, the Project Grant ``Noninvasive Sensing Measurement based on Terahertz Technology" from Province and MOE Collaborative Innovation Centre for New Generation Information Networking and Terminals, the  Key  Research  Program of CQUPT on Interdisciplinary and emerging field (A2018-01), and the venture \& Innovation Support program for Chongqing Overseas Returnees Year 2022.

\end{document}